\documentclass[prl,floatfix,twocolumn,showpacs,amsmath,amssymb
]{revtex4}
\usepackage{natbib}
\usepackage{epsfig}
\usepackage{hyperref}
\usepackage{color}
\newcommand{\unit}[1]{\ensuremath{\operatorname{#1}}}
\newcommand{\mycite}[1]{~\cite{#1}}

\newcommand{\new}[1]{#1}

\begin{document}

\title{Normal-mode spectroscopy of a single bound atom-cavity system}

\author{P.~Maunz, T.~Puppe, I.~Schuster, N.~Syassen, P.W.H.~Pinkse, and G.~Rempe}
\affiliation{Max-Planck-Institut f\"ur Quantenoptik,  Hans-Kopfermann-Str.~1,
D-85748 Garching, Germany}
\date{\today}

\begin{abstract}
  The energy-level structure of a single atom strongly coupled to the mode of
  a high-finesse optical cavity is investigated. The atom is stored in an
  intracavity dipole trap and cavity cooling is used to compensate for
  inevitable heating. Two well-resolved normal modes are observed both
  in the cavity transmission and the trap lifetime. The experiment is in good
  agreement with a Monte Carlo simulation, demonstrating our ability to
  localize the atom to within $\lambda/10$ at a cavity antinode.
\end{abstract}
\maketitle
Experimental research in quantum information science with atoms and
ions\mycite{MonroeNature02} is based on the ability to control individual
particles in a truly deterministic manner. While spectacular advances have
recently been achieved with trapped ions interacting via
phonons\mycite{RiebeNature04,BarrettNature04}, the precise control of the
motion of atoms exchanging photons inside an optical
cavity\mycite{MuenstermannPRL00} or emitting single photons on
demand\mycite{Kuhn02,McKeever04} is still a challenge. Although
very successful, experiments in cavity quantum electrodynamics with single
laser-cooled atoms\mycite{MabuchiOL96,MuenstermannOptComm99,Sauer03} are
complicated by the motion of the atom in the standing-wave mode of the optical
cavity\mycite{PinkseNature00,HoodScience00}. The lack of control over the
atomic motion is mainly due to the heating effects of the various laser fields
employed to trap and excite the atom inside the cavity in combination with the
limited ability to cool the atom between two highly reflecting mirrors facing
each other at a microscopic distance\mycite{YePRL99,FischerPRL02}.  Only
recently, good localization of the atom at an antinode of the cavity mode has
been achieved by applying optical molasses\mycite{McKeever03} or a novel
cavity cooling force\mycite{Maunz04} to a trapped atom.

In this Letter, we go one step further and employ cavity cooling to probe the
energy spectrum of a single trapped atom
\new{strongly coupled to a high-finesse
  resonator\mycite{SanchezPRL83,AgarwalPRL84}. In previous experiments using
  thermal beams, the spectrum was explored only for many
  atoms\mycite{ZhuPRL90,GrippPRA97}, one atom on
  average\mycite{ThompsonPRL92,ChildsPRL96}, or single cold atoms transiting the
  cavity\mycite{HoodPRL98}.}
Our experiment is the first in which the normal-mode (or vacuum-Rabi)
splitting of a \new{single} atom trapped inside a cavity is observed.
\new{Both the cavity transmission and the trapping time are investigated.} The
results \new{agree with a Monte Carlo simulation and} demonstrate that
remarkably good control can be obtained over this fundamental quantum system.

\begin{figure}\begin{center}
\epsfig{file=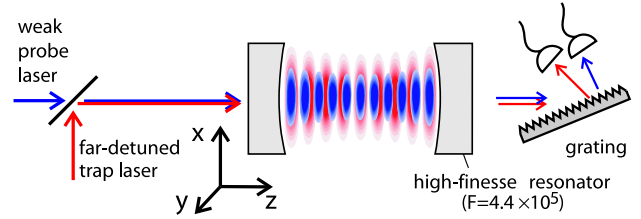,width=0.95\columnwidth}
\caption{Experimental set-up: The high-finesse cavity is excited by a weak 
  near-resonant probe field and a strong far-red-detuned trap field.
  $^{85}$Rb atoms are injected from below. Behind the cavity, the two light
  fields are separated by a grating and measured with independent
  photodetectors. }\label{Setup}
\end{center}\end{figure}

The cavity used in the experiment (Fig.~\ref{Setup}) has a finesse ${\cal
  F}=4.4\times 10^5$, a mode waist $w_0=29\unit{\mu m}$ and a length
$l=122\unit{\mu m}$\mycite{Maunz04}. A single TEM$_{00}$ mode of the
cavity is near-resonant with the $5^2S_{1/2}F=3,m_F=3 \leftrightarrow
5^2P_{3/2}F=4,m_F=4$ transition of $^{85}$Rb at $\lambda=780.2\unit{nm}$. The
atom-cavity coupling at an antinode of the standing wave,
$g/2\pi=16\unit{MHz}$, is large compared to the amplitude decay rates of the
atomic excitation, $\gamma/2\pi=3\unit{MHz}$, and the cavity field,
$\kappa/2\pi=1.4\unit{MHz}$. Strong coupling is reached, resulting in critical
photon and atom numbers $n_0=\gamma^2/2g^2 \approx 1/60$ and
$N_0=2\gamma\kappa/g^2 \approx 1/30$, respectively. This strongly coupled atom-cavity
system is probed by a weak near-resonant beam impinging on the cavity. The
probe beam is also used to cool the axial motion of the atom. A second
TEM$_{00}$ mode supported by the cavity, two free spectral ranges red detuned
with respect to the near-resonant mode, is used to trap the atom in the
cavity. This mode is resonantly excited by a trap laser at $785.3\unit{nm}$.
The far-detuned light is generated by a grating- and current-stabilized laser
diode and has a linewidth of about $20\unit{kHz}$ r.m.s.. The cavity length is
continuously stabilized to this trap laser.
The two light fields transmitted through the cavity are separated by a
holographic grating. The trap light is directed to a photomultiplier,
whereas the probe light is further filtered by a narrow-band interference
filter and then detected by two single-photon counting modules. The set-up
achieves a quantum efficiency of 32\% for the probe light transmitted through
the cavity and a suppression of the trap light on the photon counting
modules of more than $70\unit{dB}$.

Laser-cooled $^{85}$Rb atoms are injected from below by means of an atomic
fountain\mycite{MuenstermannOptComm99}. The parameters of the fountain are
chosen to get well-separated signals of single atoms which have a velocity
below $10\unit{cm/s}$. The atoms are guided into the antinodes of the
far-detuned field by a weak dipole potential with a trap depth of
$400\unit{\mu K}$. The near-resonant light used to detect the atom is blue
detuned with respect to the atomic resonance,
$\Delta_a=\omega_p-\omega_a=2\pi\times 35\unit{MHz}$, and resonant with the
cavity, $\Delta_c=\omega_p-\omega_c=0$. The presence of the atom inside the
cavity tunes the atom-cavity system out of resonance with the probe laser.
The resulting dramatic drop of the transmission allows \new{the detection of}
an atom with high signal-to-noise ratio and high bandwidth. Since the atoms
are guided into the antinodes of the far-detuned field, only atoms which enter
near the cavity center, where the antinodes of the two light fields coincide,
are strongly coupled to the probe beam and cause a deep drop of the
transmission.  Upon detection of a strongly coupled atom in the cavity, the
trap depth of the conservative dipole potential is increased to values between
$1.3$ and $1.9\unit{mK}$. This compensates for the kinetic energy of the atom
and leads to trapping. Noteworthy, all atoms which activated the trigger are
captured in the trap. We estimate the probability to trap more than one atom
at a time to be below $0.4\%$.

The storage time of a single atom in the far-detuned dipole trap without any
near-resonant light is about $30\unit{ms}$ as described in Ref.\mycite{Maunz04}. The
storage time is limited by axial parametric heating due to intensity
fluctuations of the intracavity dipole trap.
The dipole force of the probe light, which caused a shift and a distortion of
the measured spectra in earlier
experiments\mycite{MuenstermannPRL00,HoodPRL98}, can be neglected because it
is much weaker than the dipole force of the far-detuned light. However,
depending on the relative frequencies of the atomic transition, cavity resonance
and probe laser, non-conservative forces can heat or cool the
atom\mycite{HechenblaiknerPRA98,MuenstermannPRL99,MurrJPB03} mainly along the
cavity axis. In order to measure the atom-cavity spectrum, it is necessary to
probe the system at detunings for which these forces lead to strong heating.
This quickly reduces the atomic localization, and severely limits the
available probe time by boiling the atom out of the trap. To compensate the
disastrous effect of heating, cooling intervals are applied to reestablish
strong coupling of the atom to the cavity. This can be achieved by switching
the probe laser to parameters for which the velocity-dependent forces lead to
efficient cooling\mycite{Maunz04}. Of course, in the radial direction, the
atom is heated by scattering photons of the near-resonant probe light. Since
there is no radial cooling mechanism, this heating mechanism contributes to
the experimentally observed loss rate of atoms from the trap.

These considerations lead to the following protocol to perform the atom-cavity
spectroscopy: After capturing the atom in the trap, a $500\unit{\mu s}$ long
cooling interval is used to improve the localization of the atom and to
determine its coupling strength by monitoring the cavity transmission with a
resonant probe laser ($\Delta_c=0$). This is followed by a $100\unit{\mu s}$
long probe interval, where the frequency of the probe laser is changed to an
adjustable but fixed value $\Delta_c$. This sequence of cooling and probing
intervals is then repeated. As long as the atom is stored in the trap, the
transmission during the cooling intervals is low, while it is high if the atom
has left. The end of the last cooling interval during which the transmission
is below 80\% of the empty-cavity transmission determines the exit time of the
atom. Within this sequence, each probe interval is enclosed by two cooling
intervals in which the coupling strength before and after the probe interval
can be determined independently of the probing. This allows \new{the exclusion
  of} probe intervals during which the atom is only weakly coupled to the
cavity mode. \new{We find that in about 25\% of the probe intervals in which
  an atom resides in the trap, both cooling intervals have a transmission
  below 2\% of that of the empty cavity.} These probe intervals are defined as
``strongly coupled'' and are used for further analysis. The whole protocol is
repeated for different atoms and different values of $\Delta_c$.

\begin{figure}[b]\begin{center}
\epsfig{file=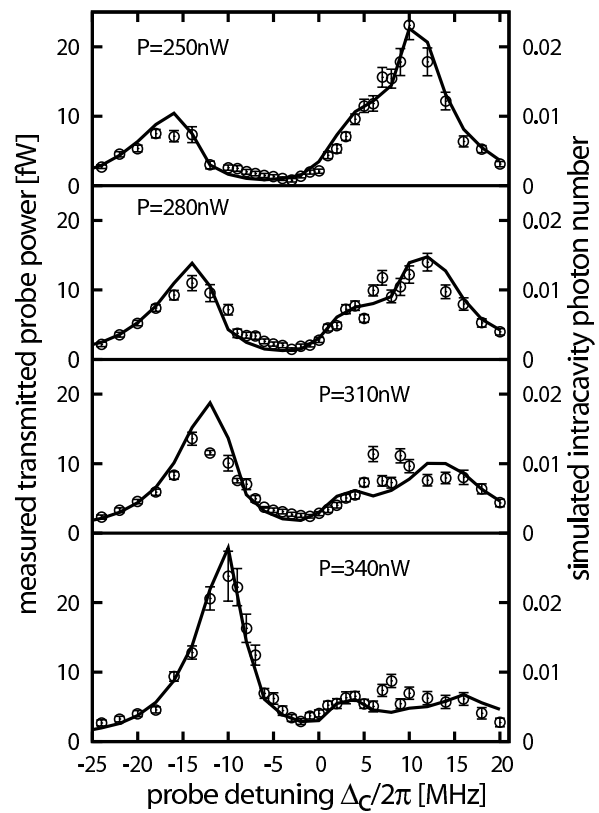,width=0.9\columnwidth}
\caption{Transmission of the cavity containing a single trapped and 
  strongly-coupled atom (circles). The detuning between the cavity and the
  atom is adjusted by tuning the Stark shift of the atom via the
  trapping-field power expressed in terms of the transmitted power, $P$. The
  average transmission during probe intervals for which the atom is found to
  be strongly coupled by independent qualification (see text) shows
  well-resolved normal-mode peaks. On average each point includes the data
  from about 350 probe intervals collected from between $35$ and $1000$ atoms.
  A Monte Carlo simulation (solid lines) describes the data
  well.}\label{SpectrumPhotons}
\end{center}\end{figure}

\new{Fig.~\ref{SpectrumPhotons} shows the average cavity transmission during
  the strongly-coupled probe intervals as a function of the probe detuning.
  The four spectra are obtained for different atom-cavity detunings and all
  show two well-resolved normal modes. Together, they display the avoided
  crossing between the atomic and the cavity resonance\mycite{MeystreSargent}.
  The atom-cavity detuning is adjusted by tuning the atomic resonance via the
  dynamic Stark effect induced by the far-detuned trap light. The induced
  (position dependent) shift, $\Delta_S$, of the atomic resonance frequency is
  proportional to the trap depth. For a transmitted power of the trap light of
  about $280\unit{nW}$ the dynamic Stark shift compensates the initial
  atom-cavity detuning of $2\pi\times 35\unit{MHz}$. The eigenstates of the
  atom-cavity system (dressed states) are superpositions of atomic ground
  state together with a cavity photon and the atomic excited state without a
  cavity photon. Since the probe laser only excites the cavity mode, the
  excitation of a dressed state is proportional to the contribution of the
  cavity state to the dressed state. This contribution depends on the
  atom-cavity detuning and explains the observation that the height of the
  left normal-mode peak increases with increasing Stark shift, while that of
  the right peak decreases. For zero detuning between atom and cavity (about
  $P=280\unit{nW}$), the contributions from the atomic and the cavity state
  are equal so that the normal modes have the same height and reach a minimum
  splitting of $2g$.  Here, the observed splitting of about $2\times
  2\pi\times 12\unit{MHz}$ is only slightly smaller than the maximal possible
  splitting of $2\times 2\pi \times 16\unit{MHz}$ expected for a point-like
  atom at rest at an antinode. This proves that the atom is localized in the
  regime of strong coupling with $g\gg (\gamma,\kappa)$.}


For a stationary atom, the widths of the two normal modes are given by a
weighted mean of the atomic and cavity linewidths. However, since the atom is
not fixed at an antinode of the probe field, but oscillates in the trap, the
atom-cavity coupling is time dependent. This leads to fluctuating frequencies
of the normal modes and therefore the measured spectra are broadened. 

The different widths of the normal modes of the spectra in
Fig.~\ref{SpectrumPhotons} can be explained by taking into account the
position-dependent Stark shift for a moving atom: An atom close to an antinode
of the trapping field experiences a larger Stark shift, which shifts both
normal modes to larger probe detunings. Near the center of the cavity, where
the antinodes of both light fields overlap, this atom is also close to an
antinode of the probe field. Therefore its coupling to the cavity is also
larger. This increases the splitting of the normal modes. Consequently, the
frequency of the left normal mode is only weakly dependent on the atomic
position while the two effects add up for the right normal mode. This broadens
the right peak to a greater extent than the left.

\begin{figure}[htb!]\begin{center}
\epsfig{file=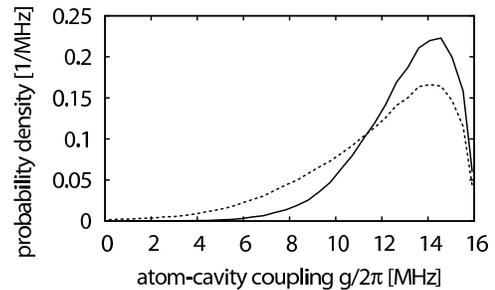,width=0.75\columnwidth}
\caption{Simulated probability distribution of the atom-cavity coupling for all
  probe intervals (dotted line) and for the strongly-coupled intervals (solid
  line). Qualification completely eliminates the occurrence of probe intervals
  with weak coupling, $g \lessapprox (\gamma,\kappa)$.}\label{figlocalisation}
\end{center}\end{figure}
The exact width and line shape of the measured normal modes are influenced by
the details of the atomic motion in the trap. Cavity heating and cooling
strongly depend on the atomic position and the frequency of the probe laser.
These forces determine the atomic motion in a complex way. In order to obtain
more information on the atomic motion we compare the measured spectra with the
results of a Monte Carlo simulation. Here a point-like atom is propagated in
space according to a stochastic differential equation for the atomic position
and momentum. The forces and momentum diffusion are given by analytic equations
for the combined atom-cavity-trap system. Parametric heating by the dipole
trap is implemented by a randomly changing potential depth. To model the
experiment in detail, single atoms are injected at random positions into the
mode. Upon activating the trigger, the trap depth of the trapping field is
increased and the atom is exposed to the alternating cooling and probing
scheme. The atomic trajectory is recorded until the atom leaves the cavity.
The simulated transmission is evaluated in the same way as the experimental
data. 
\begin{figure}[htb!]\begin{center}
\epsfig{file=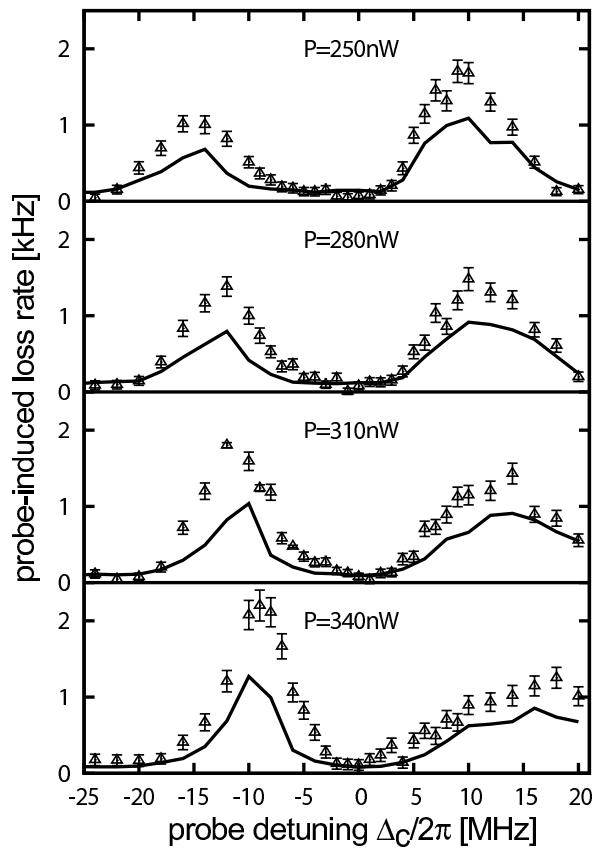,width=0.9\columnwidth}
\caption{Probe-induced loss rate of atoms from the trap (triangles) for
  different detunings between cavity and atom, adjusted by varying the
  trapping-field power. No qualification is employed. The experiment is in
  qualitative agreement with a Monte Carlo simulation (solid lines): the
  observed frequencies, widths and relative heights of the normal-mode peaks
  are well described by the simulation. Only the absolute value of the
  measured rate exceeds that of the simulation. This could be explained by
  fluctuations of experimental parameters not taken into account in the
  simulation, e.g., the lack of shot noise in the modeling of atom capture.
  For all probe detunings the simulated atomic excitation is below $1.4\%$.}
\label{SpectrumExcitation}
\end{center}\end{figure}

Results are also shown in Fig.~\ref{SpectrumPhotons} and agree well with the
experimental data if the power of the trapping field is reduced by $30\%$ with
respect to the intracavity power determined from the measured cavity
transmission. This discrepancy could be explained by different transmissions
of the two cavity mirrors. \new{For consistency, the probe light power in the
  simulation is reduced by the same amount.} The simulation also allows to
calculate the spatial probability distribution of the atom in the trap. The
axial distribution has a width (FWHM) of $\lambda/7$ if all probe intervals
are included. If only strongly-coupled probe intervals (as defined above) are
considered, the atom is axially confined to a width of $\lambda/10$ around the
antinodes of the dipole trap. The probability distribution of the simulated
atom-cavity coupling \new{(in three dimensions)} is shown in
Fig.~\ref{figlocalisation}. It shows that our selection scheme eliminates the
occurrence of probe intervals with weak atom-cavity coupling and an average
coupling of about $2\pi \times 13\unit{MHz}$ is reached. This agrees well with
the experimentally achieved coupling of $2\pi \times 12\unit{MHz}$.

Further characterization of the normal modes can be obtained by investigating
the average storage time of the atom in the trap as a function of the detuning
during the probe intervals. While the atom is probed, additional heating can
lead to a loss of the atom from the trap. \new{The probe-induced loss rate is
  shown in Fig.~\ref{SpectrumExcitation}.}  These spectra also show two
well-resolved peaks at detunings for which the excitation of the system is
high. The measurements are in \new{qualitative} agreement
with our Monte Carlo simulation. The
simulation shows that for zero and large probe detunings, spontaneous emission
accounts for about \new{$75\%$} of the probe-induced loss rate. \new{On the
  normal-mode resonances, momentum diffusion caused by dipole-force
  fluctuations of the probe light generates additional heating, which causes
  more than $80\%$ of the probe-induced loss rate. This makes the normal modes
  clearly visible in the probe-induced loss rate.}

In conclusion, cavity cooling has been applied to reliably localize a single
trapped atom in the strong-coupling region of a high-finesse cavity. Two
well-resolved normal modes are observed both in cavity transmission and the
atomic loss rate from the trap. The ability to individually excite the normal
modes of a bound atom-cavity system opens up a wealth of new possibilities
including the realization of a quantum-logic gate\mycite{TurchettePRL95} or
the control of the propagation of a light pulse\mycite{ShimizuPRL02}
with exactly one atom.


\end{document}